\documentstyle[12pt]{article}
\newlength{\mytopmargin}
\newlength{\myleftmargin}
\begin{document}

\vspace{4cm}
\noindent
{\large
{\bf JACK POLYNOMIALS AND THE
MULTI-COMPONENT  \\ CALOGERO-SUTHERLAND MODEL }}
\vspace{5mm}

\noindent
P.J.Forrester\footnote{email: matpjf@maths.mu.oz.au; supported by the ARC}\\
\noindent
Department of Mathematics, University of Melbourne,Parkville,Victoria
3052,Australia
\vspace{1cm}

\small
\begin{quote}
Using the ground state $\psi_0$ of a multicomponent generalization of the
Calogero-Sutherland model as a weight
function, orthogonal polynomials in the coordinates of one of the species are
constructed. Using evidence from exact analytic and numerical calculations,
it is conjectured that these polynomials are the Jack polynomials
$J_\kappa^{(1+1/\lambda)}$, where $\lambda$ is the coupling constant.
 The value of the normalization integral for
$\psi_0 J_\kappa^{(1+1/\lambda)}$ is conjectured, and some further related
integrals are evaluated.
\end{quote}
\normalsize
\vspace{1cm}
\noindent
{\bf 1.INTRODUCTION}
\vspace{5mm}

\noindent
The Schr\"odinger operator [1-3]
$$
{\cal H} =  - \sum_{j = 1}^N {\partial^2 \over \partial x_j^2}
+ {2 \pi^2 \over L^2}  \sum_{1 \le j < k \le N}{\lambda(\lambda - M_{jk})
\over \sin^2 \pi (x_k - x_j)/L},
\eqno (1.1)
$$
where $M_{jk}$ is the operator which exchanges the internal
`spin' coordinates of particles
$j$ and $k$, is a multi-component generalization of the periodic
$1/r^2$ many body Schr\"odinger operator originally considered by Sutherland
[4]. In fact if we seek eigenfunctions of ${\cal H}$ which are symmetric
under exchange of spin coordinates,  $M_{jk}$ can be replaced by
unity and (1.1) is precisely the Sutherland Hamiltonian. Then it has been
proved [5] that the complete set of eigenfunctions are
$$
\psi_{q,\kappa}(z_1, \dots, z_N) := \prod_{1 \le j < k \le N}
|z_k - z_j|^\lambda \prod_{j=1}^N z_j^{-q} \,
J_\kappa^{(1/\lambda)}(z_1, \dots,z_N), \quad {z_j := e^{2 \pi i x_j/L}}
\eqno (1.2)
$$
where $\kappa = (\kappa_1, \dots, \kappa_N)$,
$(\kappa_1 \ge \kappa_2 \ge \dots \ge \kappa_N \ge 0)$ is a
partition, $q \in Z_{\ge 0}$ and $J_\kappa^{(1/\lambda)}$ denotes the
Jack polynomial.

The eigenfunctions (1.2) occur in two integration formulas [6-8] of fundamental
importance in the calculation of correlations functions [9] (see also [10,11])
$$
\left ( \prod_{l=1}^N \int_{-L/2}^{L/2} d x_l \, z_l^{ (a-b)/2}
|1+z_l|^{a+b} \right )\psi_{0,0}^*(z_1, \dots, z_N)
\psi_{0,\kappa}(z_1, \dots, z_N)
$$
$$
= L^N J_\kappa^{(1/\lambda)}(1^N) \prod_{j=1}^N{\Gamma(1+a+b+\lambda(j-1))
\Gamma(\lambda j + 1) \over \Gamma(1+a+\lambda(N-j)+\kappa_j)
\Gamma(1+b+\lambda(j-1)-\kappa_j)\Gamma(1+\lambda)}
\eqno (1.3)
$$
and
$$
\left ( \prod_{l=1}^N \int_{-L/2}^{L/2} d x_l \right )
|\psi_{q,\kappa}(z_1, \dots, z_N)|^2
 = L^N{\Gamma (\lambda N + 1) \over (\Gamma
 ( \lambda + 1))^N} {(J_\kappa^{(1/\lambda)}(1^N))^2 \over
\bar{f}_N^\lambda(\kappa) f_N^\lambda(\kappa)},
\eqno (1.4)
$$
where
$$
f_n^\lambda(\kappa) := \prod_{1 \le i < j \le n}
{((j-i)\lambda + \kappa_i - \kappa_j)_\lambda \over
((j-i)\lambda)_\lambda }, \quad
\bar{f}_n^\lambda(\kappa) := \prod_{1 \le i < j \le n}
{(1 - \lambda + (j-i)\lambda + \kappa_i - \kappa_j)_\lambda \over
(1 - \lambda + (j-i)\lambda)_\lambda },
$$
with
$$
(a)_\lambda := {\Gamma(a + \lambda) \over \Gamma(a)}.
$$
The * in (1.3) denotes complex conjugation.
The explicit value of $J_\kappa^{(1/\lambda)}(1^N)$ is given by [12]
$$
(1/\lambda)^{|\kappa|}\prod_{j=1}^N {\Gamma (\lambda N - \lambda (j-1)
+ \kappa_j) \over \Gamma (\lambda N - \lambda (j-1))}.
$$

Kato and Kuramoto [3] have proved that if $M_{jk}$ in (1.1) acts as a
position coordinate exhange operator, then the ground state
 wave function $\psi_0$ which is
 symmetric in the variables
$x_1, \dots, x_{N_0}$ and antisymmetric in the variables
$x_j^{(\alpha)} := x_{N_0 + \dots + N_{\alpha - 1} + j}$ $(j = 1, \dots,
N_\alpha)$ for each $\alpha =1, \dots, p$ ( $N_0 + \dots +
N_p = N$) is given by
\renewcommand{\theequation}{1.5}
\begin{eqnarray}
\lefteqn{\psi_0(\{z_j^{(\alpha)}
\}_{\alpha = 1,\dots,p \atop j=1,\dots,N_\alpha},
\{w_j\}_{j=1,\dots,N_0})
\prod_{\alpha=1}^p \prod_{j=1}^{N_\alpha}z_j^{(\alpha)(N_\alpha - 1)/2} }
\nonumber \\
& = & \prod_{\alpha=1}^p \prod_{1 \le j < k \le N_{\alpha}}
|z_j^{(\alpha)} - z_k^{(\alpha)}|^{\lambda }(z_j^{(\alpha)} - z_k^{(\alpha)})
\prod_{1 \le j' < k' \le N_0}|w_{k'}-w_{j'}|^{\lambda}\nonumber \\
& &\times \prod_{\alpha=1}^p\prod_{\beta=1 \atop \beta \not= \alpha}^p
\prod_{j=1}^{N_\alpha}\prod_{k=1}^{N_\beta}
|z_j^{(\alpha)} - z_k^{(\beta)}|^{ \lambda }\:
\prod_{\alpha=1}^p\prod_{j=1}^{N_\alpha}\prod_{j'=1}^{N_0}
|z_j^{(\alpha)} - w_{j'}|^{ \lambda },
\end{eqnarray}
where $w_j := e^{2 \pi i x_j / L}$ and $z_j^{(\alpha)} :=
 e^{2 \pi i x_j^{(\alpha)} / L}$.

In a previous work [13], the present author has considered the analogue of
the integration formula (1.3) in the case $\kappa =0$, with $|\psi_{0,0}|^2$
replaced by the modulus squared of (1.5):
\renewcommand{\theequation}{1.6}
\begin{eqnarray}\lefteqn{D_p(N_1, \dots,N_p;N_0;a,b,\lambda)}
\nonumber \\& &
= \left ( \prod_{l=1}^{N_0} \int_{-1/2}^{1/2} d x_l \, w_l^{ (a-b)/2}
|1+w_l|^{a+b} \right )
\left (\prod_{\alpha = 1}^p \prod_{l=1}^{N_\alpha}
 \int_{-1/2}^{1/2} d x_l^{(\alpha)} \, z_l^{(\alpha) (a-b)/2}
|1+z_l^{(\alpha)}|^{a+b} \right )
 \nonumber \\
& & \times|\psi_0(\{z_j^{(\alpha)}
\}_{\alpha = 1,\dots,p \atop j=1,\dots,N_\alpha},
\{w_j\}_{j=1,\dots,N_0})|^2
\end{eqnarray}
 (for convenience we have
set $L=1$.). Guided
by data from exact numerical integration, some analytic results and the
structure of (1.3) in
the case $\kappa = 0$, we were led to conjecture the functional property
$$
D_p(N_1, \dots,N_p;N_0;a,b,\lambda)  =  f_{p-1}
(N_1,\dots,N_{p-1};N_0;
a,b,\lambda) A_p(N_1, \dots, N_p; N_0;a,b,\lambda)
\eqno (1.7a)
$$
where
\renewcommand{\theequation}{1.7b}
\begin{eqnarray}\lefteqn{A_p(N_1, \dots, N_p; N_0;a,b,\lambda)} \nonumber \\
&  & =
\prod_{j=0}^{N_p - 1} {(j+1) \Gamma ((\lambda + 1)j+a+b+\lambda
\sum_{j=0}^{p-1}
 N_j+1)
\Gamma((\lambda + 1)(j+1)+\lambda \sum_{j=0}^{p-1}N_j) \over
\Gamma(1 + \lambda)  \Gamma ((\lambda + 1)j+a+\lambda \sum_{j=0}^{p-1}N_j +1)
\Gamma ((\lambda + 1)j+b+\lambda\sum_{j=0}^{p-1}N_j +1)} \nonumber \\
\end{eqnarray}
$f_{p-1}$ can be specified by a recurrence and $N_p \ge N_j - 1 \, (j=1, \dots,
p-1)$
(a proof was given in the case $p=1$, $\lambda = 1$). This functional
property can be used as a recurrence and solved
((1.3) with $\kappa = 0$ provides the initial condition) to give $D_p$ in
terms of products of gamma functions.

The objective of this paper is to continue the study of [13] by extending the
`ground state integrand' in (1.5) to a state analogous to (1.2), and to
develop some of the integration properties of these states. These tasks are
addressed in Sections 2 and 3 respectively. In Section 4 comments and
conclusions are made, and in the Appendix a further integration formula is
noted.

\vspace{1cm}

\noindent
{\bf 2. OCCURENCE OF THE JACK POLYNOMIALS}

\vspace{.5cm}
\noindent
{\bf 2.1 The approach of Kato and Kuramoto}

\noindent
The most natural generalization of (1.2) to the multicomponent case would
be to provide the complete set of eigenfunctions of (1.1), subject to some
specification of the symmetry of the coordinates. However, since there are
both spin and position coordinates, this does not appear tractable. Another
approach is to follow
Kato and Kuramoto [3]
and to consider (1.1) as describing a spinless model with $M_{jk}$ acting as
the exchange operator for the position coordinates. Then it is shown in [3]
that seeking eigenfunctions of (1.1) of
the form
$$
\psi = \prod_{1 \le j < k \le N}
|z_k - z_j|^\lambda  \, \Phi(z_1, \dots, z_N)
\eqno (2.1)
$$
gives the eigenvalue equation for $ \Phi$
\renewcommand{\theequation}{2.2}
\begin{eqnarray}\lefteqn{
{\cal H}_N' \Phi := \bigg [ \sum_{j=1}^N
\Big (z_j {\partial \over \partial z_j} \Big )^2 + \lambda (N-1)
\sum_{j=1}^N z_j {\partial \over \partial z_j}
+ 2 \lambda \sum_{1 \le j < k \le N}{z_j z_k \over z_j - z_k}}\hspace{3.5cm}
\nonumber \\
& & \times \Big [\Big ({\partial \over \partial z_j} -{\partial \over
\partial z_k}
\Big ) - {1 - M_{jk} \over z_j - z_k}\Big ] \bigg ] \Phi = \epsilon \Phi
\end{eqnarray}
Note that for symmetric $\Phi$, $1 - M_{jk} = 0$. The analytic solutions of the
resulting eigenfunction equation are then the Jack polynomials [12].

To compute eigenfunctions for the multicomponent system, let
$\sigma$ be a partition of length $\le N$ and $P$ a permutation of
$\{1, \dots, N \}$, and put
$$
\phi_{\sigma, P} = \prod_{j=1}^N z_{P(j)}^{\sigma_j}.
\eqno (2.3)
$$
{}From [3] we know
$$
{\cal H}_N'\phi_{\sigma, P} = \sum_{j=1}^N \Big [ \sigma_j^2 + \lambda (N-1)
\sigma_j \Big ]\phi_{\sigma,P} + 2\lambda \sum_{j < k} h_{P(j) P(k)}
\phi_{\sigma, P}
\eqno (2.4a)
$$
where
$$
 h_{P(j) P(k)} = -\sigma_k + \left \{ \begin{array}{l}
\sum\nolimits_{l=1}^{\sigma_j - \sigma_k - 1}(\sigma_j - \sigma_k - l)
(z_{P(j)}^{-1}z_{P(k)})^l, \quad \sigma_j \ge \sigma_{k} + 2 \\
0, \qquad {\rm otherwise}\end{array}\right .
\eqno (2.4b)
$$
 Suppose in particular we seek a solution of (2.2) of the form
$$
\Phi = {\cal A}^{(\alpha)} \Big ( \prod_{\alpha = 1}^p \prod_{l=1}^{N_\alpha}
z_l^{N_\alpha - l} \Big ) f(w_1, \dots, w_{N_0})
\eqno (2.5a)
$$
where the symbol ${\cal A}^{(\alpha)}$ denotes the antisymmetrization in each
of the sets of variables $\{ z_1^{(\alpha)}, \dots,  z_{N_\alpha}^{(\alpha)}
\}$ $\alpha = 1, \dots, p$ and
$$
f(w_1, \dots, w_{N_0}) = \prod_{j=1}^{N_0} w_j^{\kappa_j} + \sum_{\rho <
\kappa} a_\rho \prod_{j=1}^{N_0} w_j^{\rho_j}
\eqno (2.5b)
$$
with $\kappa = (\kappa_1, \dots, \kappa_{N_0})$ a partition such that
$$
\kappa_1 \le {\rm min} (N_1, \dots, N_p)
\eqno (2.5c)
$$
($N_1, \dots, N_p$ are assumed non-zero).
 Inspection of (2.4) shows that in the action of
${\cal H}_N'$ on $\Phi$ all terms in the sum over pairs which couple
different species vanish and thus
$$
{\cal H}_N' \Phi = \sum_{\alpha = 0}^p {\cal H}_{N_\alpha}' \Phi
\eqno (2.6)
$$
where ${\cal H}_{N_\alpha}'$ is defined as in (2.2) except that the coordinates
are $w_1, \dots, w_{N_0}$ for $\alpha = 0$ and $ z_j^{(\alpha)}$ for
$\alpha = 1, \dots, p$. The condition (2.5c) is essential for the validity of
(2.6).

As noted in [3],
$\Phi$ is an antisymmetric eigenfunction of ${\cal H}_{N_\alpha}'$ for each
$\alpha = 1, \dots, p$ (this is true independent of the particular function
$f$). But we also know that the symmetric eigenfunction of ${\cal H}_{N_0}'$
of the form (2.5b) is $J_\kappa^{(1/\lambda)}(w_1, \dots,w_{N_0})$. Thus
choosing
$f = J_\kappa^{(1/\lambda)}$ in (2.5a) with $\kappa_1$ restricted as in (2.5c)
gives a class of eigenfunctions of ${\cal H}_{N}'$. Substituting this result in
(2.1)
and performing the antisymmetrizations explicitly using the Vandermonde
formula gives
\renewcommand{\theequation}{2.7}
\begin{eqnarray}
\lefteqn{\psi_\kappa(\{z_j^{(\alpha)}
\}_{\alpha = 1,\dots,p \atop j=1,\dots,N_\alpha},
\{w_j\}_{j=1,\dots,N_0})}
\nonumber \\& &
=\psi_0(\{z_j^{(\alpha)}
\}_{\alpha = 1,\dots,p \atop j=1,\dots,N_\alpha},
\{w_j\}_{j=1,\dots,N_0})
\prod_{\alpha=1}^p \prod_{j=1}^{N_\alpha}z_j^{(\alpha)(N_\alpha - 1)/2}
J_\kappa^{(1/\lambda)}(w_1,\dots,w_{N_0}).
\end{eqnarray}

To proceed further and study integration formulas, it becomes apparent that
 the states (2.7) are not orthogonal. This transpires because the operator
(1.1), with $M_{jk}$ regarded as the position coordinate exchange operator,
is not Hermitian with respect to the inner product
$$
\langle f|g \rangle_I :=
  \prod_{l=1}^{N_0} \int_{-1/2}^{1/2} d x_l \,
\prod_{\alpha = 1}^p \prod_{l=1}^{N_\alpha}
 \int_{-1/2}^{1/2} d x_l^{(\alpha)} \, f^* \, g.
\eqno (2.8)
$$
\vspace{.5cm}

\noindent
{\bf 2.2 A Gram-Schmidt approach}

\noindent
In view of the above difficulty with the  approach of Kato and Kuramoto, we
abandoned all reference to the operator (1.1) and sought to replace the
eigenstates (1.2) by the corresponding orthogonal polynomials constructed by
the Gram-Schmidt procedure.

\vspace{.2cm}
\noindent
{\bf Definition 2.1} \quad
Let $\kappa$ denote a partition of length $\le N_0$. Define a symmetric
polynomial
in the variables $w_1, \dots, w_{N_0}$, denoted $p_\kappa(w_1, \dots,
w_{N_0})$,
by the following properties:

(i) $p_\kappa(w_1, \dots, w_{N_0}) = m_\kappa + \sum_{\mu < \kappa} a_\mu
m_\mu$,
where $|\mu| = |\kappa|$, $\mu < \kappa$ is with respect to reverse
lexicographical ordering of
the partitions, $m_\mu$ refers to the monomial symmetric function with
exponents $\mu$ in the variables
$w_1, \dots, w_{N_0}$ and $a_\mu$ is the corresponding coefficient.

(ii) For all $N_1, \dots, N_p \ge \kappa_1$, $\langle \psi_0 p_\kappa
| \psi_0 p_\sigma \rangle_I = 0$ for $\kappa \ne \sigma$.

\vspace{.2cm}

\noindent
We remark that the restriction $N_1, \dots, N_p \ge \kappa_1$ is motivated
by the condition (2.5c) necessary for (2.7) to be eigenfunctions of (1.1).

The existence of these polynomials is assured by the Gram-Schmidt procedure
applied to $\{ m_\kappa \}$, where for each value of $|\kappa|$ the polynomials
$p_\kappa$ are constructed from $m_\kappa$ in reverse lexicographical ordering.
Note in particular that
$$
p_{1^k}(w_1, \dots, w_{N_0}) = m_{1^k} := \sum_{1 \le j_1 < \dots < j_k \le
N_0}
w_{j_1} w_{j_2} \dots w_{j_k}.
\eqno (2.9)
$$

\vspace{.5cm}
\noindent
{\bf 2.1.1 Exact specification of $p_{21^k}$ for $p = \lambda = 1$}

\noindent
The practical application of the Gram-Schmidt procedure requires the
computation
of multidimensional integrals, which appear intractable in general. An
exception is the case $p = \lambda = 1$, for which we have previously presented
[14] an integration procedure based on determinants to compute the
normalization of (1.5). Generalizing this procedure, we can compute,
for example, the
inner products
$$
\langle \psi_0 m_{1^{k+2}} | \psi_0 m_{2 1^k} \rangle_I \quad {\rm and} \quad
\langle \psi_0 m_{1^{k+2}} | \psi_0 m_{1^{k+2}} \rangle_I
$$
necessary to compute $a_{1^{k+2}}$ in the Definition 2.1 of $p_{21^k}$.

\vspace{.3cm}
\noindent
 {\bf Proposition 2.1} \quad
For $p = \lambda = 1$ we have
$$
p_{21^k}(w_1, \dots, w_{N_0}) =
 m_{2 1^k} + {(k+1)(k+2) \over (k+3)} m_{1^{k+2}}.
$$

\vspace{.3cm}
\noindent
 {\bf Proof} \quad
{}From (2.9) and property (i) of Definition 2.1
$$
p_{21^k} = m_{21^k} + a_{1^{k+2}} p_{1^{k+2}}.
$$
Forming the inner product with $p_{1^{k+2}}$ on both sides and using property
(ii) of Definition 2.1 gives
$$
a_{1^{k+2}} = - {\langle \psi_0 m_{1^{k+2}} | \psi_0 m_{2 1^k}
\rangle_I \over
\langle \psi_0 p_{1^{k+2}} | \psi_0 p_{1^{k+2}} \rangle_I},
\eqno (2.10)
$$
so our task is to compute these two inner products.

Let us consider the denominator of (2.10). For $p = \lambda = 1$
$$
| \psi_0(\{ z_j \}_{j=1, \dots, N_1};\{ w_j \}_{j=1, \dots, N_0} |^2
= (-1)^{N_0 + N_1} \prod_{k=1}^{N_0} w_k^{N_1} \prod_{\alpha = 1}^{N_1}
z_\alpha^{N_0 + 2N_1 - 2} A_1 A_2
\eqno (2.11a)
$$
where
$$
A_1 = \prod_{1 \le j < k \le N_0}(w_k - w_j)
\eqno (2.11b)
$$
$$
A_2 = \prod_{1 \le j < k \le N_0}(w_k^{-1} - w_j^{-1})
\prod_{j=1}^{N_0} \prod_{\alpha = 1}^{N_1} (w_j^{-1} - z_\alpha^{-1})^2
\prod_{1 \le \alpha < \beta \le N_1} (z_\beta^{-1} - z_{\alpha}^{-1})^4,
\eqno (2.11c)
$$
and furthermore, from (2.9)
$$
 p_{1^{k+2}} =  m_{1^{k+2}} =  s_{1^{k+2}} = \sum_{j_1 < \dots < j_{k+2}}
\prod_{j=1}^{k+2}
w_{j_n},
\eqno (2.12)
$$
where $s_\kappa$ denotes the Schur polynomial. Thus we can write
$$
\langle \psi_0 p_{1^{k+2}} | \psi_0 p_{1^{k+2}} \rangle_I = (-1)^{N_0 + N_1}
 \prod_{l=1}^{N_0} \int_0^1 dx_l \, w_l^{N_0}
 \prod_{\alpha=1}^{N_1} \int_0^1 dx_\alpha^{(1)} \,
z_\alpha^{N_0 + 2N_1 -2}
$$
$$
\times  s_{1^{k+2}}^*
  A_1 A_2 s_{1^{k+2}}.
\eqno (2.13)
$$

We proceed as in [14] and use a confluent form of the Vandermonde identity:
$$
 A_2 =
 \left [ \begin{array}{l} {[}w_j^{-(l-1)}]_{j =1, \dots, N_0 \atop
l =1, \dots, N_0 + 2N_1 } \\
\left [ { z_j^{-(l-1)} \atop (l-1)z_j^{-(l-2)}} \right ]_{j =1, \dots, N_1
\atop
l =1, \dots, N_0 + 2N_1 } \end{array} \right ],
\eqno (2.14)
$$
as well as the determinant formula for the Schur polynomials:
$$
A_1 s_{1^{k+2}} = \det [ w_j^{l + \kappa_{N_0 - j +1} - 1}
]_{j,l = 1, \dots, N_0}
\eqno (2.15)
$$
where $\kappa_j = 1$ $(j=1, \dots, k+2)$, $\kappa_j = 0$ otherwise.

Since (2.14) and (2.15) are antisymmetric with respect to interchanges
of $w_1, \dots, w_{N_0}$, in the integral (2.13) we can replace
(2.15) by $N_0!$ times its diagonal term. Multiply each factor
$w_j^{j+ \kappa_{N_0 - j + 1} - 1}$ of
this term into the row of (2.14) dependent on $w_j$. Now take the sum
in the definition (2.12) of $s_{1^{k+2}}^*$ outside the integral and
multiply the factors of the summand into appropriate rows of (2.14).
Row-by-row integration of the determinant with respect to
$w_1, \dots, w_{N_0}$ gives a non-zero contribution in row $2N_1 + j$
only in column
$$
l = N_1 + j + \kappa_{N_0 - j + 1} - \xi_j, \qquad
\xi := \left \{ {1 \quad {\rm if} \quad j=j_1, \dots,j_n \atop 0,
\quad {\rm otherwise}}
\right .
\eqno (2.16)
$$
and this term is equal to unity.

For the determinant to be non-zero we require each of the columns (2.16) to
be distinct. Thus we require
$$
\xi_1 = \dots = \xi_\nu = 1, \quad \xi_{N_0 - k -1}= \dots = \xi_{N_0 - \nu}
= 1, \quad \xi_j = 0 \: \: {\rm otherwise}
\eqno (2.17)
$$
for some $\nu = 0,\dots ,k+2$ or equivalently
$$
\{j_1, \dots, j_k \} =\{1, \dots, \nu, N_0 - k -1, \dots, N_0  - 2  -\nu \}.
\eqno (2.18)
$$
Assuming (2.17) and expanding the integrated determinant by the
columns (2.16) with non-zero entries, gives, after expanding the remaining
terms
and grouping in pairs
\renewcommand{\theequation}{2.19a}
\begin{eqnarray}\lefteqn{
\langle \psi_0 p_{1^{k+2}} | \psi_0 p_{1^{k+2}} \rangle_I}\nonumber \\ & & =
N_0! \sum_{P(2 \alpha) > P(2\alpha - 1)} \epsilon (P) \prod_{\alpha = 1}^{N_1}
(P(2 \alpha) - P(2 \alpha - 1)) \int_{-1/2}^{1/2} dx_\alpha^{(1)} \,
z_\alpha^{N_0 + 2N_1 + 1 - P(2 \alpha) - P(2 \alpha - 1)}\nonumber \\
\end{eqnarray}
where
$$
P(\alpha) \in \{ 1, \dots, N_1-1\} \cup \{N_1 + \nu \}
\cup \{ N_1 + N_0 - \nu +1 \} \cup \{ N_1 + N_0 +2, \dots, N_0 + 2N_1 \}
\eqno (2.19b)
$$

We see that it is possible to choose
$$
P(2 \alpha - 1) = N_0 + 2N_1 + 1 - P(2 \alpha), \quad \alpha = 1, \dots, N_1
\eqno (2.20a)
$$
and thus have a non-zero contribution to (2.19a) if and only if
$$
P(2 \alpha ) \in \{ N_1 + N_0 +1 - \nu \} \cup \{ N_1 + N_0 +2, \dots, 2N_1 +
N_0
\}
\eqno (2.20b)
$$
Each of the $N_1!$ different choices (2.20b) give the same contribution to
(2.19a) and so
\renewcommand{\theequation}{2.21}
\begin{eqnarray}
\langle \psi_0 p_{1^{k+2}} | \psi_0 p_{1^{k+2}} \rangle_I & = &
N_0! N_1! \prod_{l =1}^{N_1 - 1}(N_0 + 2N_1 + 1 - 2l) \sum_{\nu = 0}^{k+2}
(N_0 + 1 - 2 \nu) \nonumber \\
& = & N_0! N_1!(k+3)(N_0 -1 -k) \prod_{l =1}^{N_1 - 1}(N_0 + 2N_1 + 1 - 2l)
\end{eqnarray}

A similar calculation gives
$$
\langle \psi_0 p_{21^{k}} | \psi_0 p_{1^{k+2}} \rangle_I  =
- N_0! N_1!(k+1)(k+2)(N_0 -1 -k) \prod_{l =1}^{N_1 - 1}(N_0 + 2N_1 + 1 - 2l)
\eqno (2.22)
$$
The stated result follows by substituting (2.21) and (2.22) in (2.10).

\vspace{.4cm}
For $\lambda \in Z_{\ge 0}$ we noted in [13]  a simple method of exact
numerical integration for integrals of
the type in (2.10), which relies on the fact that the integrand can be written
as a finite (multivariable) Fourier series. Applying this method to
compute (2.10) for $\lambda = 2,3$, $N_1=1$ and various (small) values of $k$
and $N_0$ indicates that Proposition 2.1 has a simple generalization to the
general $\lambda$ case

\vspace{.3cm}
\noindent
 {\bf Conjecture 2.1}\quad
For $p=1$ and $\lambda \ge 0$,
$$
p_{21^k}(w_1, \dots, w_{N_0}) =
 m_{2 1^k} + {\lambda(k+1)(k+2) \over (\lambda(k+2) + 1)} m_{1^{k+2}}.
$$

\vspace{.3cm}
A striking feature of Proposition 2.1 and more generally Conjecture 2.1 is that
the coefficients are independent of both $N_0$ and $N_1$. Exact numerical
evaluation of $p_\kappa$ for other $\kappa$ indicate that this is a general
property, which is valid provided $N_1 \ge \kappa_1 - 1$.

\vspace{.3cm}
\noindent
 {\bf Conjecture 2.2} \quad
For general $p \in Z^+$ and $\lambda \ge 0$ the coefficients $a_\nu$ in the
polynomials of Definition 2.1 are independent of $N_0, N_1, \dots, N_p$. Also,
in (ii) of Definition 2.1 the condition $N_1,\dots,N_p \ge \kappa_1$ can be
replaced by
$N_1, \dots,N_p \ge \kappa_1 - 1$.

\vspace{.3cm}
The independence  of the coefficients on $N_1, \dots, N_p$ can be understood in
terms of the following conjecture which generalizes the recurrence (1.7a).

\vspace{.2cm}
\noindent
{\bf Conjecture 2.3}\quad
Let $h(w_1, \dots, w_{N_0})$ be a Laurent polynomial of the form
$$
h = \sum_{\sigma \le \rho} c_\sigma m_\sigma,
$$
where $\rho = (\rho_1, \dots, \rho_{N_0})$,
$|\rho_1| \ge \dots \ge |\rho_{N_0}|$. Suppose $N_1 \ge |\rho_1|$ and
$N_1 \le \dots \le N_p$. Then
\begin{eqnarray*}\lefteqn{\langle \psi_0 |  \psi_0
\prod_{l=1}^{N_0} w_l^{(a-b)/2}|1 + w_l|^{a+b}
\prod_{\alpha = 1}^p \prod_{l=1}^{N_\alpha}
z_l^{(\alpha)(a-b)/2} |1 + z_l^{(\alpha)}|^{a+b} h \rangle}\\
& & =   f_{p-1}^{ \rho}
(N_1,\dots,N_{p-1};N_0;
a,b,\lambda) A_p(N_1, \dots, N_p; N_0;a,b,\lambda)
\end{eqnarray*}
where $A_p$ is given by (1.7b) and $f_{p-1}^{ \rho}$ satisfies the recurrence
\begin{eqnarray*}\lefteqn{f_{k-1}^\rho(N_1, \dots, N_{k-1};N_0;a,b,\lambda)}
 \\
& & = A_{k-1}(N_1, \dots, N_{k-1}; N_0;a,b,\lambda)
f_{k-2}^\rho(N_1, \dots, N_{k-2};N_0;a,b,\lambda)
\end{eqnarray*}
where
$$
A_{k-1}(N_1, \dots, N_{k-1}; N_0;a,b,\lambda)
:= \prod_{j=1}^{N_{k-1}}{A_k(N_1, \dots,N_{k-2},j-1,j;N_0;a,b,\lambda)
\over A_k(N_1, \dots,N_{k-2},j,j-1;N_0;a,b,\lambda)}
$$
with
$
f_0^\rho(N_0;a,b,\lambda)
$ equal to the inner product in the case $p=0$.

\vspace{.3cm}
The recurrence for $f_{p-1}^\rho$ is analogous to the recurrence for
$f_{p-1}$ in (1.7b), which is given in ref.~[13]. Taken in the order
$k=p,p-1,\dots,2$ these equations explicitly determine
$f_{p-1}^\rho$ and thus give the inner product in terms of the inner product
in the case $p=0$. Now,
the ratio of inner products specifying each $a_\nu$ in Definition 2.1 are of
the same form as the inner product in Conjecture 2.3 with $a=b=0$. By
introducing the recurrence in Conjecture 2.3 we conclude that the
dependence on $N_1, \dots, N_p$ factorizes from the dependence on $\kappa$
and $N_0$ and thus cancels out of the ratios.

By definition (see e.g.~[12]), the Jack polynomials satisfy property (i) of
Definition 2.1 with the coefficients $a_\nu$ independent of $N_0$. Let
us compare $p_{21^k}$, as given by Conjecture 2.1, with
$J_{21^k}^{(\alpha)}$, which is given by [12, Proposition 7.2 with the
normalization chosen so that the coefficient of $m_{21^k}$ is unity]:
$$
J_{21^k}^{(\alpha)}(w_1, \dots,w_{N_0}) = m_{21^k} +
{(k+1)(k+2) \over k + 1 + \alpha} m_{1^{k+2}}.
$$
We see
$$
p_{21^k} = J_{21^k}^{(1 + 1/\lambda)}
$$
which together with our exact numerical data suggests that the polynomials
$p_\kappa$ are simply related to the Jack polynomials.

\vspace{.2cm}
\noindent
{\bf Conjecture 2.4} \quad For general $\lambda \ge 0$ and partitions
$\kappa$, the polynomials $p_\kappa$ of Definition 2.1 are given in
terms of the Jack polynomials by
$$
p_\kappa(w_1, \dots,w_{N_0}) = J_\kappa^{(1 + 1/\lambda)}
(w_1, \dots,w_{N_0}).
$$

\vspace{.5cm}
 \noindent
{\bf 3. NORMALIZATION INTEGRAL}

\vspace{.2cm}
In this Section we will provide the (in general conjectured) exact evaluation
of the normalization integral, which is given by the inner product
$$
\langle \psi_0 J_\kappa^{(1 + 1/\lambda)}(w_1, \dots,w_{N_0})|
 \psi_0 J_\kappa^{(1 + 1/\lambda)}(w_1, \dots,w_{N_0}) \rangle_I
=: {\cal N}_p^\kappa(N_1, \dots,N_p;N_0;\lambda)
\eqno (3.1)
$$
where $\psi_0$ is given by (1.5) and the inner product by
(2.8), and the condition (2.5c) on $\kappa_1$ is assumed.

According to Conjecture 2.3 it suffices to consider the case $p=1$ and
$N_1 = \kappa_1$. Note that in the case $p=1$ Conjecture 2.3 gives
$$
{\cal N}_1^\kappa(N_1;N_0;\lambda)=
{\cal N}_1^\kappa(\kappa_1;N_0;\lambda)
\prod_{j = \kappa_1}^{N_1-1} {(j+1) \over \Gamma (1 +\lambda)}
((\lambda + 1)j + \lambda N_0 + 1)_\lambda
\eqno (3.2)
$$

\vspace{.2cm}
\noindent
{\bf 3.1 Analytic result for $\lambda = p = 1$, $\kappa = 1^k$}

\noindent
In the case $\lambda = p = 1$, $\kappa = 1^k$, analytic results can be obtained
by reading off a result obtained in the proof of Proposition 2.1.

\vspace{.2cm}
\noindent
{\bf Proposition 3.1} \quad Consider the case  $\lambda = p = 1$, $\kappa =
1^k$
of (3.1). For $N_1 \ge 1$ we have
\begin{eqnarray*}{\cal N}_1^\kappa(N_1;N_0;\lambda) & = &
N_0! N_1! (k+1)(N_0 +1 -k) \prod_{l=1}^{N_1 - 1} (N_0+2N_1 +1 -2l) \\
& = & N_0!  (k+1)(N_0 +1 -k) \prod_{l=1}^{N_1 - 1}(l+1)(2l+N_0+1).
\end{eqnarray*}

\vspace{.2cm}
\noindent
{\bf Proof} \quad The first line is precisely (2.21) with $k$ replaced by
$k-2$. The second line, which is of the general form (3.2), follows after
replacing $l$ by $N_1 - l$ in the product and noting
 $N_1! = \prod_{l=1}^{N_1 - 1}(l+1)$.

\vspace{.3cm}
\noindent
{\bf 3.2 The conjecture}

\noindent
Let us write the partition $\kappa = (\kappa_1, \dots, \kappa_{N_0})$ as
$$
\prod_{j=1}^{\kappa_1} (\kappa_1 + 1 - j)^{f_{\kappa_1 + 1 -j}}
\eqno (3.3)
$$
so that $f_j$ gives the frequency of the integer $j$ in the partition, and
denote
$$
{\cal N}_1^\kappa(N_1;N_0;\lambda) =:
{\cal N}_1^{f_{\kappa_1} f_{\kappa_1 -1}\dots f_1}(N_1;N_0;\lambda)
\eqno (3.4)
$$

We seek to evaluate (3.4) with $N_1=\kappa_1$
in terms of the variables $f_{\kappa_1},
\dots, f_1$. There are some general properties that our expression must
possess. First, from (3.2), we require
$$
{\cal N}_1^{(f_{\kappa_1} = 0) f_{\kappa_1 -1}\dots f_1}(\kappa_1;N_0;\lambda)
= {\cal N}_1^{ f_{\kappa_1 -1}\dots f_1}(\kappa_1 - 1;N_0;\lambda)
 {\kappa_1 \over \Gamma (1 + \lambda)} ((\lambda + 1)(\kappa_1 - 1)
+ \lambda N_0 + 1 )_\lambda
\eqno (3.5)
$$

Second, in the cases $\kappa_{N_0} > 0$, since [12]
$$
 J_\kappa^{(1 + 1/\lambda)}(w_1, \dots,w_{N_0}) =
\prod_{j=1}^{N_0} w_j^{\kappa_{N_0}} \,
J_{\kappa - \kappa_{N_0}}^{(1 + 1/\lambda)}(w_1, \dots,w_{N_0}),
$$
where $\kappa-\kappa_{N_0} = (\kappa_1 - \kappa_{N_0}, \dots, \kappa_{N_0}-
\kappa_{N_0})$, we require
$$
{\cal N}_1^\kappa(N_1;N_0;\lambda) = {\cal N}_1^{\kappa - \kappa_{N_0}}
(N_1;N_0;\lambda).
\eqno (3.6)
$$

Guided by Proposition 3.1, the general properties (3.5) and (3.6), and some
exact numerical data, we were led to formulate the following conjecture.

\vspace{.2cm}
\noindent
{\bf Conjecture 3.1} \quad Define ${\cal N}_1^\kappa(N_1;N_0;\lambda)$ by
(3.1) and introduce the notation (3.3) and (3.4). Then
\begin{eqnarray*}
\lefteqn{{\cal N}_1^{f_{\kappa_1} f_{\kappa_1 -1}\dots
f_1}(\kappa_1;N_0;\lambda)
={\kappa_1! \Gamma(\lambda N_0 + 1) \over
(\Gamma(1+\lambda))^{N_0 + \kappa_1}}
 \prod_{j=1}^{\kappa_1}\Big (\lambda f_j + 1 \Big )_\lambda  }\\
& &
 \times
 \prod_{j=1}^{\kappa_1}{\Big ((\lambda + 1)j + 1 + \lambda
\sum_{k=1}^{\kappa_1 + 1 -j} f_{j+k-1} \Big )_\lambda
\Big ((\lambda + 1)(j-1) + 1 + \lambda
(N_0 - \sum_{k=1}^{\kappa_1 + 1 -j} f_{j+k-1}) \Big )_\lambda \over
\Big ( (\lambda + 1)j + 1 + \lambda f_j \Big )_\lambda}
\end{eqnarray*}

\vspace{.2cm}
\noindent
{\bf 3.2 Evaluation of some multidimensional integrals}

\noindent
Knowledge of the normalization allows some multidimensional integrals
represented by inner products to be evaluated. Suppose
$f(w_1, \dots, w_{N_0})$ is a polynomial of degree $K$ and has a known
expansion in terms of Jack polynomials:
$$
f(w_1, \dots, w_{N_0}) = \sum_{k=0}^K \sum_{|\sigma| = k}
c_\sigma J_\sigma^{(1 + 1/\lambda)}(w_1, \dots, w_{N_0})
\eqno (3.7)
$$
Multiplying both sides by $|\psi_0|^2
J_\kappa^{(1 + 1/\lambda)}(w_1^*, \dots, w_{N_0}^*)$, where $\psi_0$ is given
by (1.5) with $N_1 \ge K$, and integrating using the orthogonality property of
Definition 2.1  gives
$$
\langle \psi_0 J_\sigma^{(1 + 1/\lambda)}(w_1, \dots, w_{N_0}) |
\psi_0 f(w_1, \dots, w_{N_0}) \rangle = c_\kappa
{\cal N}_p^\kappa(N_1, \dots,N_p;N_0;\lambda).
\eqno (3.8)
$$
Assuming Conjecture 3.1, the multidimensional integral defining the inner
product
is thus evaluated.

Perhaps the most straightforward example of an application of (3.8) is given
by the following result.

\vspace{.2cm}
\noindent
{\bf Proposition 3.2} \quad Suppose $N_1 \ge 1$ and put $N= N_0 + N_1$.
We have
\begin{eqnarray*}\lefteqn{
  \prod_{l=1}^{N} \int_{-1/2}^{1/2} d x_l \, \prod_{1 \le j < k \le N}
 |e^{2 \pi i x_k} - e^{2 \pi i x_j}|^{2 \lambda} } \nonumber \\& &
\times
 \prod_{N_0+1 \le j < k \le N} |e^{2 \pi i x_k} - e^{2 \pi i x_j}|^2
\prod_{l=1}^{N_0}(1 + u e^{2 \pi i x_l})
J_{1^k}^{(1 + 1/\lambda)}(e^{-2 \pi i x_1}, \dots, e^{-2 \pi i x_{N_0}})\\
& & = u^k {\Gamma(\lambda N_0 + 1) \over (\Gamma(1 + \lambda))^{N_0 + 1}}
(\lambda k + 1)_\lambda (\lambda (N_0 - k) + 1)_\lambda
\prod_{l=1}^{N_1 - 1} {(l+1) \over \Gamma(1+\lambda)}((\lambda + 1) l + \lambda
N_0 + 1)_\lambda
\end{eqnarray*}

\vspace{.2cm}
\noindent
{\bf Proof} \quad This follows immediately from the simple expansion formula
$$
\prod_{l=1}^{N_0}(1 + u w_l) = \sum_{k=0}^{N_0} u^k m_{1^k} = \sum_{k=0}^{N_0}
u^k  J_{1^k}^{(1 + 1/\lambda)}(w_1, \dots, w_{N_0}),
\eqno (3.9)
$$
(3.8) and Conjecture 3.1 with $\kappa_1 = 1$ and $f_{\kappa_1} = k$.

\vspace{.3cm}
In the Appendix we give an integration formula which replaces
$$
\prod_{l=1}^{N_0}(1 + u e^{2 \pi i x_l}) \quad {\rm by} \quad
\prod_{l=1}^N e^{\pi i x_l (a-b)} |1 + e^{2 \pi i x_l} |^{a+b}
$$
in Proposition 3.2, and thus can be viewed as a generalization of (1.3) in
the case $\kappa = 1^k$.

\vspace{.5cm}
\noindent
{\bf 4.  CONCLUSION}

\vspace{.2cm}
\noindent
Our objective of extending the ground state integrand in (1.5) to a state
analogous to (1.2) led us to define (Definition 2.1)) a class of
multivariable symmetric polynomials. As stated in Conjecture 2.4, it appears
that these polynomials are precisely the Jack polynomials
$J_\kappa^{1+1/\lambda}$. An immediate
difficulty in proving this result is the lack of a differential operator
defining the states as eigenfunctions. In particular, the operator (1.1)
with $M_{jk}$ interpreted as acting on the particle position coordinates
does not have these states as eigenfunctions, and in fact is not an
Hermitian operator.

The states we defined appear to possess closed form,
product-of-gamma-function-type expressions for their
 normalization (Conjecture 3.1). However they
do not form a complete set as they are restricted to the
coordinates $w_1, \dots, w_{N_0}$. Even for functions of the
coordinates $w_1, \dots, w_{N_0}$ they do not form a complete set, due to the
restriction $N_1 \ge \kappa_1$. In particular, this means that the
results obtained so far  are not sufficient to compute correlation functions
in multicomponent Calogero-Sutherland-type systems.

\vspace{.5cm}
\noindent
{\bf ACKNOWLEDGEMENTS}

\vspace{.2cm}
\noindent
I thank the referee for a careful reading.

\vspace{.5cm}
\noindent
{\bf APPENDIX}
\vspace{.2cm}

\noindent
Here we will evaluate the integral in Proposition 3.2, modified so that
$$
\prod_{l=1}^{N_0}(1 + u e^{2 \pi i x_l}) \quad \mbox{is replaced by} \quad
\prod_{l=1}^N e^{\pi i x_l (a-b)} |1 + e^{2 \pi i x_l} |^{a+b}
$$
in the integrand. Thus we are seeking to evaluate
\renewcommand{\theequation}{A.1}
\begin{eqnarray}\lefteqn{
D_1^{1^k}(N_1;N_0;a,b,\lambda):=
 \left ( \prod_{l=1}^{N} \int_{-1/2}^{1/2} d x_l \, e^{\pi i x_l (a-b)}
|1+e^{2 \pi i x_l}|^{a+b} \right )} \nonumber \\& &
\times \prod_{1 \le j < k \le N} |e^{2 \pi i x_k} - e^{2 \pi i x_j}|^{2
\lambda}
 \prod_{N_0+1 \le j < k \le N} |e^{2 \pi i x_k} - e^{2 \pi i x_j}|^2
J_{1^k}^{(1 + 1/\lambda)}(e^{2 \pi i x_1}, \dots, e^{2 \pi i x_{N_0}})
\nonumber \\
\end{eqnarray}

According to Conjecture 2.3
$$
D_1^{1^k}(N_1;N_0;a,b,\lambda) = D_1^{1^k}(1;N_0;a,b,\lambda)
{A_1(N_1;N_0;a,b,\lambda) \over A_1(1;N_0;a,b,\lambda)},
\eqno ({\rm A}.2)
$$
where $A_1$ is given by (1.7b). Since $J_{1^k}^{(1 + 1/\lambda)} = m_{1^k}$,
we see that
$$
D_1^{1^k}(1;N_0;a,b,\lambda)=
{N_0 + 1 - k \over N_0 + 1} D_0^{1^k}(N_0+1;a,b,\lambda)
\eqno ({\rm A}.3)
$$
which is an explicit evaluation, as the
value of $ D_0^{1^k}$ is given explicitly by (1.3) (in this formula the
Jack polynomial is assumed to have the normalization of Stanley [12]; to
normalize $J_{1^k}^{(1 + 1/\lambda)}$ as in Definition 2.1 it is necessary to
divide by $k!$).  Thus
\renewcommand{\theequation}{A.4}
\begin{eqnarray}\lefteqn{
D_1^{1^k}(N_1;N_0;a,b,\lambda)=
{(-N_0)_k (b/\lambda)_k \over k!(-N_0 - (a+1)/\lambda)_k}
D_0^0(N_0 + 1;a,b,\lambda)}   \nonumber \\& &
\times \prod_{j=1}^{N_1 - 1} {(j+1) \Gamma ((\lambda + 1)j+a+b+\lambda
 N_0+1)
\Gamma((\lambda + 1)(j+1)+\lambda N_0) \over
\Gamma(1 + \lambda)  \Gamma ((\lambda + 1)j+a+\lambda N_0 +1)
\Gamma ((\lambda + 1)j+b+\lambda N_0 +1)}
\end{eqnarray}
($D_0^0$ is given by (1.3) with $\kappa = 0$, $N = N_0 +1$ and $L=1$).

{}From (A.4) a generalization of the Selberg integral of the type first given
by Aomoto [15] can be deduced. Thus using (3.9)
we see from (3.1) that
\renewcommand{\theequation}{A.5}
\begin{eqnarray}\lefteqn{
\sum_{k=0}^{N_0} u^k D_1^{1^k}(N_1;N_0;a,b,\lambda)=
 \left ( \prod_{l=1}^{N} \int_{-1/2}^{1/2} d x_l \, e^{\pi i x_l (a-b)}
|1+e^{2 \pi i x_l}|^{a+b} \right )} \nonumber \\& &
\times \prod_{1 \le j < k \le N} |e^{2 \pi i x_k} - e^{2 \pi i x_j}|^{2
\lambda}
 \prod_{N_0+1 \le j < k \le N} |e^{2 \pi i x_k} - e^{2 \pi i x_j}|^2
\prod_{l=1}^{N_0}(1 + u e^{2 \pi i x_l}),
\end{eqnarray}
where $N = N_0 + N_1$ and $N_1 \ge 1$. On the other hand, from (A.4) we have
\renewcommand{\theequation}{A.6}
\begin{eqnarray}\lefteqn{
\sum_{k=0}^{N_0} u^k D_1^{1^k}(N_1;N_0;a,b,\lambda)} \nonumber \\& &
= {}_2F_1 ( -N_0, b/\lambda; -N_0 - (a+b)/\lambda; u) D_0^0(N_0 +
1;a,b,\lambda)
 \nonumber \\& &
\times \prod_{j=1}^{N_1 - 1} {(j+1) \Gamma ((\lambda + 1)j+a+b+\lambda
 N_0+1)
\Gamma((\lambda + 1)(j+1)+\lambda N_0) \over
\Gamma(1 + \lambda)  \Gamma ((\lambda + 1)j+a+\lambda N_0 +1)
\Gamma ((\lambda + 1)j+b+\lambda N_0 +1)},
\end{eqnarray}
which says the integral in (A.5) is proportional to a certain Gauss
hypergeometric
function, with proportionality constant given in terms of products of gamma
functions.

\vspace{1cm}

\noindent
{\bf REFERENCES}

\begin{description}
\item[1] A.P. Polychronakos, Phys. Rev. Lett. {\bf 69}, (1992) 703.
\item[2] Z.N.C. Ha and F.D.M. Haldane, Phys. Rev. Lett. B {\bf 46} (1992)
9359
\item[3] Y. Kato and Y. Kuramoto, Phys. Rev. Lett. {\bf 74} (1995) 1222
\item[4] B. Sutherland, Phys. Rev. A {\bf 4} (1971) 2019
\item[5] P.J. Forrester, Nucl. Phys. B {\bf 416} (1994) 377
\item[6] I.G. Macdonald, {\it Hall polynomials and symmetric functions} 2nd
ed. (Oxford Univ. Press, Oxford, 1995)
\item[7] K.W.J. Kadell, Compos. Math. {\bf 87} (1993) 5
\item[8] J. Kaneko, SIAM J. of Math. Analysis {\bf 24} (1993) 1086
\item[9] P.J. Forrester, Mod. Phys. Lett. {\bf 9} (1995) 359
\item[10] Z.N.C. Ha, Nucl. Phys. B {\bf 435} (1995) 604
\item[11] F. Lesage, V. Pasquier and D. Serban, Nucl. Phys. B {\bf 435}
(1995) 585
\item[12] R.P. Stanley, Adv. Math. {\bf 77} (1989) 1243
\item[13] P.J. Forrester, Int. J. of Mod. Phys. B {\bf 9} (1995) 1243
\item[14] P.J. Forrester and B. Jancovici, J. Physique Lett. {\bf 45} (1984)
L583
\item[15] K. Aomoto, SIAM J. of Math. Analysis {\bf 18} (1987) 545
\end{description}

\end{document}